\documentclass[final,3p,times]{elsarticle}

\makeatletter
\def\ps@pprintTitle{%
 \let\@oddhead\@empty
 \let\@evenhead\@empty
 \let\@oddfoot\@empty
 \let\@evenfoot\@empty
}
\makeatother

\usepackage[utf8]{inputenc}
\usepackage{tikz}
\usetikzlibrary{matrix}
\usepackage{bm}

\usepackage{amsmath}

\journal{}

\usepackage{amsmath}
\usepackage{amsfonts,color}
\usepackage{amsmath}
\usepackage{amsthm}
\usepackage{pdfpages}
\usepackage{amsmath}
\usepackage{empheq}
\usepackage{amsfonts,color}
\usepackage{amssymb,float}
\usepackage{physics}
\usepackage{booktabs,multirow}
\usepackage{titlesec}

\usepackage{hyphenat}
\usepackage{ragged2e}

\titleformat{\paragraph}
{\normalfont\normalsize\bfseries}{\theparagraph}{1em}{}
\titlespacing*{\paragraph}
{0pt}{3.25ex plus 1ex minus .2ex}{1.5ex plus .2ex}

\usepackage{amsmath}
\usepackage[utf8]{inputenc}
\allowdisplaybreaks
\usepackage{color}
\usepackage{hyperref}
\hypersetup{
    colorlinks=true,
    linkcolor=blue,
    filecolor=magenta,      
   citecolor=blue
}

\usepackage{enumitem}

\usepackage{tikz}
\usetikzlibrary{shapes.geometric}
\usetikzlibrary{arrows.meta,arrows}

\begin{document}

\begin{frontmatter}
\title{An exact five dimensional Weyl-Geometry Gauss-Bonnet Black Hole}

\author[1,2]{Sebastian Bahamonde}
\ead{sbahamondebeltran@gmail.com}
\author[3]{ Máximo Bañados}
\ead{mbanados@uc.cl}

\address[1]{Kavli Institute for the Physics and Mathematics of the Universe (WPI), The University of Tokyo Institutes
for Advanced Study (UTIAS), The University of Tokyo, Kashiwa, Chiba 277-8583, Japan.}
\address[2]{Cosmology, Gravity, and Astroparticle Physics Group, Center for Theoretical Physics of the Universe,
Institute for Basic Science (IBS), Daejeon, 34126, Korea.}
\address[3]{Facultad de Física, Pontificia Universidad Católica de Chile, Avenida Vicuña Mackema 4860, Santiago, Chile.}

\begin{abstract}
We present a new exact black hole solution of a 5-dimensional Weyl-geometry Gauss-Bonnet theory of gravity. The Euclidean sector defines a fully regular metric coupled to the Weyl vector field. The Euclidean action and entropy are computed, with the latter following the simple $A/4$ form plus a term linear in the horizon radius, characteristic of Gauss-Bonnet couplings.
\end{abstract}

\end{frontmatter}

\section{Introduction}\label{sec:intro}

It is well known that the Gauss-Bonnet invariant is a topological term that does not modify the field equations in a 4-dimensional theory. However, in higher dimensions, this term, which is quadratic in curvature, is no longer a boundary term and contributes to the field equations. Black holes in higher-dimensional Gauss-Bonnet gravity were found by Boulware and Deser~\cite{Boulware:1985wk}. These solutions generalize the Schwarzschild-de Sitter metric in higher dimensions, with the Gauss-Bonnet term appearing as a correction to the geometry of spacetime~\cite{Wheeler:1985nh}. The thermodynamics of the Boulware-Deser black holes has been analyzed by several authors, including~\cite{Myers:1988ze,Jacobson:1993xs,Cai:2001dz,Clunan:2004tb}. In 5 dimensions, a new phase of locally stable small black holes emerges when the Gauss-Bonnet coefficient is below a critical threshold. Beyond this critical value, the black holes remain thermodynamically stable under all conditions~\cite{Cai:2001dz}. 

The Boulware-Deser black hole lives on the   Riemannian sector of the theory, meaning that both torsion and nonmetricity are assumed to be zero (as in General Relativity). In this paper, we report a new 5-dimensional solution of this type in Weyl geometry, where not only curvature is present but nonmetricity is also non-vanishing~\cite{Weyl:1918ib,Romero:2012hs}. The nonmetricity present in Weyl geometry is of a particular type, defined solely by a vector known as Weyl vector. In these geometries, the traceless part of nonmetricity is absent, implying that only the dilational part of matter sources nonmetricity. On Weyl geometry the length of vectors is not preserved under parallel transport (angles are preserved). This feature is linked to invariance under Weyl transformations of the Weyl connection. For some recent work and more details on this geometry, see~\cite{Hehl:1994ue,Haghani:2023nrm,Barcelo:2017tes,Romero:2012hs,Banados:2024rfy}.

This paper is organized as follows: In Sec.~\ref{sec:theory}, we construct a \(d\)-dimensional Einstein-Gauss-Bonnet gravity theory within Weyl geometry. Sec.~\ref{sec:BH} is devoted to studying the 5-dimensional case, where we find a new black hole solution.  In Sec \ref{thermo} we study the Euclidean sector of this theory, compute the black hole action and entropy. 
Finally, we summarize our main results in Sec.~\ref{sec:conclusions}. Quantities with tildes denote that they are computed with respect to the affine connection with nonmetricity, whereas quantities without symbols are Riemannian.

\newpage
\section{Gauss-Bonnet gravity in Weyl geometry }~\label{sec:theory}
Weyl geometry is a particular case in Metric-Affine geometry where torsion is vanishing and nonmetricity is solely characterised by its Weyl vector. This means that nonmetricity (defined as the covariant derivative of the metric) can be written as
\begin{equation}
\tilde{\nabla}_{\lambda}g_{\mu \nu}=g_{\mu\nu}W_{\lambda}\,,
\end{equation}
where $W_{\lambda}$ is so-called Weyl vector. Here, the covariant derivative is taken with respect to the general affine connection $\tilde{\Gamma}^{\lambda}\,_{\mu \nu}$ given by
\begin{equation}
\tilde{\Gamma}^{\lambda}\,_{\mu \nu}=\Gamma^{\lambda}\,_{\mu \nu}+\frac{1}{2} \Big(g_{\mu \nu } W^{\lambda }  -   \delta^{\lambda }{}_{\mu } W_{\nu }-  \delta^{\lambda }{}_{\nu } W_{\mu }\Big)\,,
\end{equation}
where $\Gamma^{\lambda}\,_{\mu \nu}$ is the Levi-Civita connection. Then, the curvature is defined as
\begin{align}\label{totalcurvature}
\tilde{R}^{\lambda}\,_{\rho \mu \nu}&=\partial_{\mu}\tilde{\Gamma}^{\lambda}\,_{\rho \nu}-\partial_{\nu}\tilde{\Gamma}^{\lambda}\,_{\rho \mu}+\tilde{\Gamma}^{\lambda}\,_{\sigma \mu}\tilde{\Gamma}^{\sigma}\,_{\rho \nu}-\tilde{\Gamma}^{\lambda}\,_{\sigma \nu}\tilde{\Gamma}^{\sigma}\,_{\rho \mu}\,,\\
&=R^{\lambda }{}_{\rho \mu \nu }+\frac{1}{2}  g_{ \rho [\mu }\delta^{\lambda }{}_{\nu] } W_{\alpha } W^{\alpha } + \frac{1}{2} W_{[\mu } g_{\nu] \rho } W^{\lambda } + \frac{1}{2} \delta^{\lambda }{}_{[\mu } W_{\nu ]} W_{\rho } +\nabla_{[\mu }W^{\lambda } g_{\nu] \rho }  -  \delta^{\lambda }{}_{\rho } \nabla_{[\mu }W_{\nu] } -   \nabla_{[\mu }W_{|\rho| }\delta^{\lambda }{}_{\nu] }\,.
\end{align}
Here, $R^{\lambda }{}_{\rho \mu \nu }$ is the Levi-Civita Riemann curvature and the covariant derivatives are related to the Levi-Civita connection. The generalized Gauss-Bonnet invariant is
\begin{equation}
\label{GG}\tilde{\mathcal{G}}=\tilde{R}_{\mu\nu\lambda\rho}\tilde{R}^{\lambda\rho\mu\nu}-(\tilde{R}_{\mu\nu}+\hat{R}_{\mu\nu})(\tilde{R}^{\nu\mu}+\hat{R}^{\nu\mu})+\tilde{R}^2\,,
\end{equation}
where we have defined the Ricci and co-Ricci tensors as
\begin{equation}
\tilde{R}_{\mu\nu}=\tilde{R}^{\lambda}\,_{\mu \lambda \nu}\,, \quad \hat{R}_{\mu\nu}=\tilde{R}_{\mu}\,^{\lambda}\,_{\nu\lambda}\,.
\end{equation}
Note that the generalized Gauss-Bonnet invariant is a boundary term in 4-dimensions (even in the presence of torsion).

Now, by performing a post-Riemannian expansion, 
\begin{equation}
\label{GGB}\tilde{\mathcal{G}}=\mathcal{G}+\mathcal{G}_2\,,
\end{equation}
where $\mathcal{G}$ is the Riemannian Gauss-Bonnet invariant
\begin{equation}
    \mathcal{G}=R^2-4R_{\mu\nu}R^{\mu\nu}+R_{\mu\nu\alpha\beta}R^{\mu\nu\alpha\beta}\,,
\end{equation}
and the extra term coming from the Weyl part of nonmetricity $\mathcal{G}_2$ becomes
\begin{align}
    \mathcal{G}_2&= 2 (3 -  d) \Big[G_{\alpha \beta } W^{\alpha } W^{\beta } + \frac{1}{4} (d-3 ) (d-2) W_{\alpha } W^{\alpha } \nabla_{\beta }W^{\beta } + 2 G_{\alpha \beta } \nabla^{\beta }W^{\alpha }\Big]\nonumber\\
    &+ (3 -  d) (d-2 ) \Big[\frac{1}{2} R W_{\alpha } W^{\alpha }+ W^{\alpha } W^{\beta } \nabla_{\beta }W_{\alpha } -  \nabla_{\alpha }W^{\alpha } \nabla_{\beta }W^{\beta } + \nabla_{\alpha }W_{\beta } \nabla^{\beta }W^{\alpha }\Big]\nonumber\\
    &+\frac{1}{16} (d-4 ) (d-3 ) (d-2 ) (d-1 ) W_{\alpha } W^{\alpha } W_{\beta } W^{\beta } \,,
\end{align}
 where $d$ is the dimension of the manifold. By ignoring boundary terms, the above relation can be simplified after integrating by parts using 
\begin{eqnarray}
   \nabla_{\lambda }W^{\lambda } \nabla_{\mu }W^{\mu }&=& (G_{\mu \nu } + \frac{1}{2} g_{\mu \nu } R) W^{\mu } W^{\nu } + \nabla_{\mu }W_{\nu } \nabla^{\nu }W^{\mu }+\textrm{b.t.}\,,\\
   W_{\alpha } W_{\beta } \nabla^{\beta }W^{\alpha }&=&- \frac{1}{2} W_{\alpha } W^{\alpha } \nabla_{\beta }W^{\beta }+\textrm{b.t.}\,,
\end{eqnarray}
yielding
\begin{align}
    \mathcal{G}_2&=(d-4 ) (d-3 ) G_{\alpha \beta } W^{\alpha } W^{\beta } + \frac{1}{16} (d-4 ) (d-3 ) (d-2 ) (d-1 ) W_{\alpha } W^{\alpha } W_{\beta } W^{\beta } \nonumber\\
& -  \frac{1}{2} (d-4 ) (d-3 ) (d-2 ) W_{\alpha } W^{\alpha } \nabla_{\beta }W^{\beta }+\textrm{b.t.}\,,\label{G2B}
\end{align}
 where we can easily notice that if $d=4$, $\mathcal{G}_2$ acts as a boundary term. 

In this letter, we consider the generalization of the $d$-dimensional Riemannian Gauss-Bonnet gravity\footnote{Note that in principle one can also modify this action by changing $R\rightarrow \tilde{R}$, but due to the identity 
\begin{eqnarray}
 \tilde{R}&=& R -  \frac{1}{4} (d-2 ) (d-1 ) W_{\mu } W^{\mu } + (d-1 ) \nabla_{\mu }W^{\mu }
\end{eqnarray} that Lagrangian would only introduce a mass to the Weyl part of nonmetricity.}:
\begin{align}
		I_{\rm GB-WC} 
		&=\frac{1}{16\pi}\int \dd^d\!x \sqrt{-g} \Big[R+\frac{(d-1)(d-2)}{l^2}+
		\alpha \tilde{\mathcal{G}}\Big]  \,.\label{scalarSTGAA}
\end{align}
where $\alpha$ measures the strength of the quadratic and Weyl contributions. We have also introduced a negative cosmological constant $\Lambda=-(d-1)(d-2)/(2l^2)$. In Weyl geometry, we consider the Lagrangian in terms of the metric and the Weyl vector. Explicitly, the Lagrangian in terms of the metric and Weyl vector reads (up to boundary terms)
\begin{align}
	I_{\rm GB-WC}[g_{\mu\nu},A_\mu] 
		&=\frac{1}{16\pi}\int \dd^d\!x \sqrt{-g} \Big[R+\frac{(d-1)(d-2)}{l^2}+
		\alpha \mathcal{G}+(d-4 ) (d-3 ) \alpha \Big(G_{\mu \nu } W^{\mu } W^{\nu } \nonumber\\
  &\,\,\,\,\,\,\,\,\,\,\,\,\,\,\,\,\,\,\,\,\,\,\,\,\,\,\,\,\,\,\,\,\,\,+ \frac{1}{16}(d-2 ) (d-1 ) W_{\mu } W^{\mu } W_{\nu } W^{\nu }-  \frac{1}{2} (d-2 ) W_{\mu } W^{\mu } \nabla_{\nu }W^{\nu }\Big)\Big]   \,.\label{scalarSTG}
\end{align} 

The associated equations of motion are,
\begin{align}
    G_{\mu\nu}&=\frac{1}{2}  \Big[\frac{(1 -  d) (2 - d) }{ l^2}+ \alpha \mathcal{G}  \Big]g_{\mu \nu }  + 2 \alpha \Big(2 R_{\mu }{}^{\alpha } R_{\nu \alpha } - R_{\mu \nu } R + 2 R^{\alpha \beta } R_{\mu \alpha \nu \beta } -R_{\mu }{}^{\alpha \beta \rho } R_{\nu \alpha \beta \rho }\Big)\nonumber\\
    &-\frac{1}{32} (3 -  d) (d-4 ) (d-2 )\beta\Big[ 4 W_{\mu } W_{\nu } \bigl((-1 + d) W_{\alpha } W^{\alpha } - 4 \nabla_{\alpha }W^{\alpha }\bigr) + g_{\mu \nu } W^{\alpha } W^{\beta } \bigl((-1 + d) W_{\alpha } W_{\beta } + 16 \nabla_{\beta }W_{\alpha }\bigr)\Big]\nonumber\\
    &+ \frac{1}{2} (d-4 ) (d-3 ) \alpha\Big[g_{\mu \nu } \bigl(- \nabla_{\alpha }W^{\alpha } \nabla_{\beta }W^{\beta } -  (\nabla_{\alpha }W_{\beta } - 2 \nabla_{\beta }W_{\alpha }) \nabla^{\beta }W^{\alpha }\bigr) + \nabla^{\alpha }W_{\nu } \nabla_{\mu }W_{\alpha } + \nabla_{\alpha }W^{\alpha } \nabla_{\mu }W_{\nu } \nonumber\\
    &+ W_{\nu } (- \nabla_{\alpha }\nabla^{\alpha }W_{\mu } + \nabla_{\mu }\nabla_{\alpha }W^{\alpha }) - 2 \nabla_{\mu }W^{\alpha } \nabla_{\nu }W_{\alpha } + \nabla^{\alpha }W_{\mu } (-2 \nabla_{\alpha }W_{\nu } + \nabla_{\nu }W_{\alpha }) + \nabla_{\alpha }W^{\alpha } \nabla_{\nu }W_{\mu } \nonumber\\
    &+ W_{\mu } (- \nabla_{\alpha }\nabla^{\alpha }W_{\nu } + \nabla_{\nu }\nabla_{\alpha }W^{\alpha }) + W^{\alpha } \Bigl(G_{\mu \nu } W_{\alpha } + 2 R_{\alpha \mu \nu \beta } W^{\beta } + G_{\alpha \nu } W_{\mu } + G_{\alpha \mu } W_{\nu } \nonumber\\
    &+ g_{\mu \nu } \bigl(R W_{\alpha } + 2 (G_{\alpha \beta } W^{\beta } -  \nabla_{\beta }\nabla_{\alpha }W^{\beta } + \nabla_{\beta }\nabla^{\beta }W_{\alpha })\bigr) + \nabla_{\mu }\nabla_{\alpha }W_{\nu } + \nabla_{\nu }\nabla_{\alpha }W_{\mu } - 2 \nabla_{\nu }\nabla_{\mu }W_{\alpha }\Bigr)\Big]\,,\label{fieldeq1}
\end{align}
while by varying it with respect to the Weyl vector we obtain
\begin{align}
    0&=(d-4 ) (d-3 )\alpha \Big[2 G_{\mu \alpha } W^{\alpha } + \frac{1}{4} (d-2 ) (d-1 )  W_{\alpha } W^{\alpha } W_{\mu } + (2 - d)  (W_{\mu } \nabla_{\alpha }W^{\alpha } -  W_{\alpha } \nabla_{\mu }W^{\alpha })\Big]\,.\label{fieldeq2}
\end{align}
In the next section we will find a  black hole solution in 5 dimensions in this theory and study its thermodynamics properties.

\section{Exact Gauss-Bonnet black hole solution with dilations }\label{sec:BH}
We now concentrate on the five dimensional case with static and spherically symmetric fields. This means, 
\begin{equation}
    \mathcal{L}_{\xi}g_{\mu\nu}=\mathcal{L}_{\xi}W_\mu=0\,.
\end{equation}
with
\begin{align}
    \xi_{(0)}&=\partial_{t}\,,\\
    \xi_{(1)}&=\cos\phi \partial_\theta-\cot\theta \sin\phi \partial_\phi\,,\\
    \xi_{(2)}&=\sin\phi \cos\psi\partial_\theta+\cot\theta\cos\phi\cos\psi\partial_\phi-\frac{\cot\theta\sin\psi}{\sin\phi}\partial_\psi\,,\\
    \xi_{(3)}&=\sin\phi \sin\psi\partial_\theta+\cot\theta\cos\phi\sin\psi\partial_\phi+\frac{\cot\theta\cos\psi}{\sin\phi}\partial_\psi\,,\\
     \xi_{(4)}&=\cos\psi\partial_\phi-\cot \phi \sin\psi \partial_\psi\,,\\
       \xi_{(5)}&=\sin\psi\partial_\phi+\cot\phi \cos\psi \partial_\psi\,,\\
     \xi_{(6)}&=\partial_  \psi\,.
\end{align}
The metric in these coordinates is:
\begin{align}\label{sph_metric}
   \dd s^2=-\Psi_0(r)\dd t^2+\frac{1}{\Psi_1(r)}\dd r^2+R(r)^2(\dd \theta^2+\sin^2 \theta\, \dd \phi^2+\sin^2 \theta \sin^2\phi\,\dd \psi^2)\,,
\end{align}
and the Weyl field (with spherical symmetry) is
\begin{eqnarray}
    W=w_0(r) \, dt + w_1(r)\, dr \label{sph_W}
\end{eqnarray}
Hereafter we will assume coordinates such that $R(r)=r$. Recall that there is no gauge symmetry (for $W_\mu$), so that, in general, $w_r$ cannot be set to zero. 

The field equations on the spherically symmetric field are the following. For the Weyl field, equations  \eqref{fieldeq2} reduce to
\begin{eqnarray}
0&=&\alpha  w_0 \Big[r^2 w_1 \Psi_1 \Psi_0'+\Psi_0 \left(\Psi_1 \left(2 r^2 w_1'-2 r^2 w_1{}^2+6 r w_1-4\right)+r \left(r w_1-2\right) \Psi_1'+4\right)+2 r^2 w_0{}^2\Big]\,,\label{connec1}\\
0&=&\alpha  \Big[2 r^2 w_0 \Psi_0 w_0'+r^2 w_0{}^2 \left(2 w_1 \Psi_0-\Psi_0'\right)+w_1 \Psi_0 \Big(\Psi_0 (4-2 \left(r^2 w_1{}^2-3 r w_1+2\right) \Psi_1)+r (r w_1-2) \Psi_1 \Psi_0'\Big)\Big]\,,\label{connec2}
\end{eqnarray}
For the metric field equations~\eqref{fieldeq1} reduce to
\begin{eqnarray}
 0&=& -\frac{2 \Psi_0 }{l^2 r^2}\Big(l^2 \left(r \Psi_1'+2 \Psi_1\right)-2 \left(l^2+2 r^2\right)\Big)+  \alpha  \Big[\frac{5 w_0{}^4}{\Psi_0}-4 w_1 w_0 \Psi_1 w_0'\nonumber\\
 &&+2 w_0{}^2 \Big\{\Psi_1 \left(2 w_1'-\frac{6}{r^2}+w_1 \left(\frac{6}{r}-3 w_1+\frac{2 \Psi_0'}{\Psi_0}\right)\right)+\frac{r \left(r w_1-3\right) \Psi_1'+6}{r^2}\Big\}\nonumber\\
 &&+\frac{\Psi_0 }{r^3}\Big\{r w_1 \Psi_1 \Big(w_1 \left(\left(r^2 w_1{}^2-4\right) \Psi_1-4\right)+4 r \left(r w_1-2\right) \Psi_1 w_1'\Big)+2 \Big((r w_1-2){}^2 (r w_1+1) \Psi_1-4\Big) \Psi_1'\Big\}\Big]\,,\label{metriceq1}\\
 0&=&2 r \Psi_0 \Big[2 \Psi_0 \left(l^2 \Psi_1-l^2-2 r^2\right)+l^2 r \Psi_1 \Psi_0'\Big]+\alpha  l^2 \Big[-r^3 w_0{}^4+2 r w_0{}^2 \Big\{\Psi_0 \Big((3 r^2 w_1{}^2+2) \Psi_1-2\Big)-r (r w_1+1) \Psi_1 \Psi_0'\Big\}\nonumber\\
 &&+\Psi_0 \Psi_1 \Big\{2 \Big((r^2 w_1{}^2 \left(r w_1-1\right)-4) \Psi_1+4\Big) \Psi_0'+r w_1{}^2 \Psi_0 \Big(12-(r w_1-2)(5 r w_1-2) \Psi_1\Big)\Big\}\nonumber\\
 &&+4 r^2 (r w_1+2) w_0 \Psi_0 \Psi_1 w_0'\Big]\,,\label{metriceq2}\\
 0&=&\Psi_0 \Big[4 \Psi_0{}^2 \Big\{l^2 \left(r \Psi_1'+\Psi_1\right)-l^2-6 r^2\Big\}+l^2 r \Psi_0 \Big\{\Psi_0' \Big(r \Psi_1'+4 \Psi_1\Big)+2 r \Psi_1 \Psi_0''\Big\}-l^2 r^2 \Psi_1 \Psi_0'{}^2\Big]\nonumber\\
 &&+\alpha  l^2 \Big[w_0{}^2 \Big\{\Psi_0{}^2 \left(\left(6 r^2 w_1{}^2+4\right) \Psi_1+4 r \Psi_1'-4\right)+3 r^2 \Psi_1 \Psi_0'{}^2+r \Psi_0 \left(-\Psi_0' \left(\left(6 r w_1+4\right) \Psi_1+r \Psi_1'\right)-2 r \Psi_1 \Psi_0''\right)\Big\}\nonumber\\
 &&+\Psi_0 \Big\{w_1{}^2 \Psi_1 \left(4 \Psi_0{}^2 \left(\Psi_1 \left(1-3 r^2 w_1'\right)+3 r \Psi_1'+1\right)-r^2 \Psi_1 \Psi_0'{}^2+r \Psi_0 \left(\Psi_0' \left(3 r \Psi_1'+4 \Psi_1\right)+2 r \Psi_1 \Psi_0''\right)\right)\nonumber\\
 &&+4 \Big(\Psi_0 \left(\Psi_1 \left(2 r^2 w_0'{}^2-3 \Psi_0' \Psi_1'-2 \left(\Psi_1-1\right) \Psi_0''\right)+\Psi_0' \Psi_1'\right)+\left(\Psi_1-1\right) \Psi_1 \Psi_0'{}^2\Big)-6 r^2 w_1{}^3 \Psi_0{}^2 \Psi_1 \Psi_1'-3 r^2 w_1{}^4 \Psi_0{}^2 \Psi_1{}^2\nonumber\\
 &&+4 r w_1 \Psi_0 \Psi_1{}^2 w_1' (r \Psi_0'+4 \Psi_0)\Big\}-3 r^2 w_0{}^4 \Psi_0+4 r w_0 \Psi_0 \Big\{w_0' \Big(\Psi_0 \left((3 r w_1+4) \Psi_1+r \Psi_1'\right)-2 r \Psi_1 \Psi_0'\Big)\nonumber\\
 &&+2 r \Psi_0 \Psi_1 w_0''\Big\}\Big]\,,\label{metriceq3}
\end{eqnarray}
where primes denote derivatives with respect to the radial coordinate $r$.

Remarkably, despite the complicated and non-linear nature of this system the exact solution is available. Taking combinations of all equations one first find,  
\begin{equation}
 \Psi_0(r)=\Psi_1(r) \equiv \Psi(r).
\end{equation}
and then by recursively solving all equations, one finds
\begin{eqnarray}
    w_0(r)&=&\epsilon \sqrt{\frac{C_1^2 r^2}{16}+\frac{\frac{C_1 \kappa}{2}+(\Psi-2) \Psi+1}{r^2}-\frac{1}{2} C_1 (\Psi+1)+\frac{\kappa^2}{r^6}+\frac{2 \kappa (\Psi-1)}{r^4}}\,,\label{w0}\\
    w_1(r)&=&\frac{C_1 r}{4 \Psi}+\frac{\kappa}{r^3 \Psi}-\frac{1}{r \Psi}+\frac{1}{r}\,,\label{wr} \\
    \Psi(r)&=& \frac{1}{1+4 \alpha  \kappa/r^4} \left(1 -\frac{8 M }{3\pi r^2}+\frac{r^2}{l^2}+\frac{4 \alpha  \kappa}{r^4}-\frac{2 \alpha  \kappa^2}{r^6}\right)\,, \label{Psi}
\end{eqnarray}
where $C_1,\kappa$ and $M$ are arbitrary integration constants. As a check, note that the metric function $\Psi(r)$ takes the usual Schwarzschild form (in 5d), with corrections vanishing when $\alpha=0$. The constant $M$ is still the total mass, as show in next section. One notices that the constant $\kappa$ appears in the Weyl vector as $\propto r^{-3}$ suggesting that this quantity is not acting like a charge since a $\propto r^{-2}$ would be needed for that in 5 dimensions. Then, this constant might be associated with a kind of new dipole-type configuration that is related to the dilations. For positive $M$ and $\alpha$, this solution describes a black hole configuration with a single horizon. It is important to note that this solution significantly differs from the Boulware-Deser black hole \cite{Boulware:1985wk,Garraffo:2008hu} (standard Riemannian geometry).\footnote{It is also possible to define another theory with the following action
\begin{align}
		\mathcal{S} 
		&=\frac{1}{16\pi}\int \dd^d\!x \sqrt{-g} \Big[R+\frac{(d-1)(d-2)}{l^2}+
		\alpha \mathcal{G}+\beta \mathcal{G}_2\Big]  \,,\label{scalarSTGAAA}
\end{align}
that is a generalization of~\eqref{scalarSTGAA} with two arbitrary constants that separate the Riemannian and the post-Riemannian sectors. By assuming spherical symmetry in 5 dimensions as in~\eqref{sph_metric}-\eqref{sph_W} we find that there is a unique exact spherically symmetric solution to the theory. This solution has the same Weyl vector as displayed in~\eqref{w0}-\eqref{wr} and the metric functions are 
\begin{eqnarray}
    \Psi_0(r)=\Psi_1(r)=\Psi(r)=1+\frac{\beta  \kappa}{r^2 (\alpha -\beta )}+\frac{r^2}{4 (\alpha -\beta )} \left(1+\sigma  \sqrt{1+\frac{16 M (\alpha -\beta )}{r^4}-\frac{8 (\alpha -\beta )}{l^2}+\frac{16 \alpha  \beta  \kappa^2}{r^8}}\right)\,,
\end{eqnarray}
where $\epsilon=\pm 1, \sigma=\pm 1$, $\kappa$ and $C_1$ are constants that are related to the Weyl vector and $M$ is an integration constant that can be interpreted as the mass of the black hole. 
The above solution is a generalization of the Lovelock black hole also known as the Boulware-Deser black hole (see~\cite{Boulware:1985wk,Garraffo:2008hu}) with $\beta$ being the constant that modifies it. Note that the case $\alpha=\beta$ that is the theory that we considered in our paper is not covered by this solution. However, we do not explore this theory in detail here, as in the context of Weyl geometry, it lacks a compelling motivation.}

\section{Action and entropy}
\label{thermo}

We now turn to the calculation of the entropy of this black hole.  This poses some challenges due to the non-standard geometry and coupling to higher order curvatures.  The result is, however, simple: the entropy contains the classic $A/ 4$ contribution plus a term linear in the horizon radius, characteristic of a Gauss-Bonnet coupling. 

Over the years, several different ways to compute the action and entropy of black holes have been devised. Starting with the work of Gibbons-Hawking \cite{Gibbons:1976ue}, the Brown-York approach \cite{Brown:1992br}, Hamiltonian methods \cite{Jacobson:1993xs,Banados:1992wn,Wald:1993nt}, and more recently, background independent methods suggested by the AdS/CFT correspondence\cite{Balasubramanian:1999re,Papadimitriou:2005ii}. Given the exotic nature of our system we will resort to the simplest approach --background subtraction-- supplemented by a finite boundary term to ensure the right variational principle. That is, we first focus on the variation of the action (see \cite{Andrade:2006pg}), adding necessary finite terms to ensure a correct variational principle when the inverse temperature $\beta$ is fixed. By ensuring that the horizon is regular, the value of the action follows straightforwardly. 

The Euclidean field reads
\begin{eqnarray}
    ds^2&=&\Psi(r)\beta^2 \dd\tau^2+\frac{1}{\Psi(r)}\dd r^2+r^2(\dd \theta^2+\sin^2 \theta\, \dd \phi^2+\sin^2 \theta \sin^2\phi\,\dd \psi^2)\,, \\
    W &=&i\,w_0(r) \beta d\tau+w_1(r)dr\,,
\end{eqnarray}
where the functions $\Psi(r),w_0(r),w_1(r)$ are given in (\ref{w0},\ref{wr},\ref{Psi}). Here we have explicitly introduced the Euclidean period $\beta$ (and fix the range of the Euclidean time coordinate to be $0\leq\tau<1$).  

This solution has 4 integration constants: $C_1,\kappa,M$ and $\beta$, with $M$ is the total mass of the system (see below). However, some restrictions on the space of parameters arise by demanding regularity of the horizon. There are two  (disconnected) branches of Euclidean solutions. 

The first branch is defined by the choice, 
\begin{eqnarray}
\kappa=0=M, \ \ \ \ \  C_1=-{4 \over l^2}\,. 
\end{eqnarray} 
The metric and vector field functions are, 
\begin{eqnarray}
w_0(r) &=& {i\over l^2} \sqrt{ 1 + {r^2 \over l^2}} \,,\\
w_1(r) &=& 0\,, \\
\Psi(r) &=& 1 + {r^2 \over l^2 }\,.
\end{eqnarray} 
This solution is regular for all $r>0$ and its only free parameter is $\beta$, that can take any value. We label this solution as the thermal AdS background.

The second branch of regular solutions is obtained by the choice of parameters, 
\begin{eqnarray}
    C_1=0\,,\quad \kappa=r_{+}^2\,, \label{conds}
\end{eqnarray}
where $r_+$ is the point defined by $\Psi(r_+)=0$, that is, the location of the horizon. This is the black hole sector. Note that this sector is disconnected from thermal AdS. In this sense, AdS appears as a bound state. 

After imposing (\ref{conds}) the functions (\ref{w0},\ref{wr},\ref{Psi}) simplify to, 
\begin{eqnarray}
    w_0(r)&=&i\frac{\left(r^2-r_+^2\right)   \left(r^4-2 \alpha  l^2\right)}{l^2 r^3 \left(r^4+4 \alpha  r_+^2\right)}\,,\\
    w_1(r)&=&\frac{\left(r^2+r_+^2\right) \left(r^4-2 \alpha  l^2\right)}{r_+^2 \left(2 \alpha  l^2 r+r^5\right)+l^2 \left(r^5-2 \alpha  r^3\right)+r^7}\,,\\
    \Psi(r)&=&\frac{\left(r^2-r_+^2\right) \left(r_+^2 \left(2 \alpha  l^2+r^4\right)+l^2 \left(r^4-2 \alpha  r^2\right)+r^6\right)}{l^2 r^2 \left(r^4+4 \alpha  r_+^2\right)}\,.
\end{eqnarray}
We see that both $w_0(r)$ and $\Psi(r)$ vanish at the point $r=r_+$. The constant $M$ appearing in (\ref{w0},\ref{wr},\ref{Psi}) has been expressed a function of $r_+$ via $\Psi(r_+)=0$. This gives,
\begin{eqnarray}
  M(r_+)=\frac{3 \pi  \alpha }{4} \left(1 + \frac{1}{2\alpha}  r_+^2 + \frac{r_+^4}{2\alpha l^2} \right)
  \label{eqM}.
\end{eqnarray}

The conditions (\ref{conds}) are necessary on the black hole sector because the Euclidean time one-form $d\tau$ is singular at the horizon. As a consequence $w_0(r)$ must vanish there. Moreover $w_0(r)$ must be analytic (the derivatives must be well-behaved there) and that imposes a second condition. Note that $C_1=0$ implies $w_0(r)=-\Psi(r)w_1(r)$ and makes the nonmetricity gauge form non-singular (in $r\neq0$) as it was already reported in~\cite{Bahamonde:2020fnq}.  A third regularity condition, the Gibbons-Hawking temperature condition, will fix $\beta$ as a function of the mass.

Despite the presence of 4 initial constants, this black hole carries no hair, only mass. All other constants are uniquely defined via regularity conditions. This is consistent with the fact that this theory does not have gauge invariance; there is no asymptotic charge associated to $W_\mu$.

Having identified the regular fields, we turn to the problem of evaluating the action on the on-shell fields. Our sign convention for the Euclidean action will be $iI_M = - I_E$ with $I_E>0$. In practice we consider the action (\ref{scalarSTGAA}) on the Euclidean field and multiplied it by -1. An efficient way to evaluate an action on-shell is to first look at its {\it variation}. Since the field can be made regular in the whole interior we know that only boundary terms at the spatial infinity are relevant. If one had considered a Hamiltonian action, with the Schwarzschild time as a quantization coordinate, a singular behaviour appears at the horizon and the entropy appears as a boundary term there. We take the Gibbons-Hawking idea by evaluating the {\it covariant} action without contributions from the horizon. Of course, both methods give the same result. 

Before analysing the variational problem, we will subtract an infinity by considering the difference of actions, 
\begin{equation}
\tilde I[g,W] = I[g,W] - I_0
\end{equation}
where $I_0$ is the action evaluated on the AdS bound state. This substraction cancels exactly a divergent term coming from the black hole action leading to $\tilde I[g,W]$ being finite. Now, we need to make sure it has the right variational properties. 
 
For the spherically symmetric field it is enough to consider a minisuperspace model with variables $\Psi_0,\Psi_1,w_0,w_r$. The action has the form (by direct replacement of the spherically symmetric field into the covariant action, we skip a rather long expression since it is not too informative), 
\begin{equation}
\tilde I[\Psi_0,\Psi_1,w_0,w_1] = \int_{r_+}^\infty dr \, L(\Psi_0,\Psi_1,w_0,w_1;\Psi_0',\Psi_1',w_0',w_1';\Psi_0'')  + B\,.   \label{Is}
\end{equation}
The terms involving the second derivative $\Psi_0''$ are a total derivative not affecting the equations of motion. Nevertheless, we will keep it because its elimination would change the boundary terms. Adding a piece $d/dr(...)$ would change the structure at $r_+$ adding a new boundary term there. We stick to the covariant action with only boundary terms at infinity. The variation of (\ref{Is}) has the form, 
\begin{eqnarray}
\delta \tilde I &=& (\textrm{eom}) + \left.\left(  {\delta L  \over \delta \Psi_0'} \delta \Psi_0 + {\delta L  \over \delta \Psi_1'} \delta \Psi_1 + {\delta L  \over \delta \Psi_0''} \delta \Psi_0' - \left( {\delta L \over \delta \Psi_0''}\right)' \delta \Psi_0  +  {\delta L  \over \delta w_0'} \delta w_0 +  {\delta L  \over \delta w_1'} \delta w_1 \right) \right|^\infty + \delta B \\
&=& -{1 \over 3}\beta\delta M + {2\pi\alpha \over l^2} \beta \delta r_+^2  + \left(  {2M \over 3} + {\pi\alpha r_+^2 \over l^2} \right)\delta\beta + \delta B\,,
\end{eqnarray}
where `\textrm{eom}' stands for equations of motion. In the second line we have replaced the on-shell field. We now see clearly the role of the asymptotic boundary term $B$: it must make sure that the variation has the form (something)$\delta\beta$. The necessary value for $B$ is then, 
\begin{equation}
B =   {1 \over 3} M \beta - {\pi \alpha \beta r_+^2  \over l^2}\,,
\end{equation}
and $\delta \tilde I$ becomes, 
\begin{equation}
\delta \tilde I = M \delta \beta \label{dI}\, ,
\end{equation}
confirming that the parameter $M$ plays the role of total mass for this solution.  

Asymptotically, $M$ and $\beta$ are independent parameters. However, there is one more regularity condition to impose: the Gibbons-Hawking period to avoid conical singularities in the interior.  This condition relates the extensive parameter $M$ to the intensive $\beta$.  In simple systems like a Schwarzschild black hole, this relation is trivial but in more complicated cases, like ours, it is often difficult to express $M(\beta)$ explicitly, albeit the relation does exist.  A useful workaround is to express all quantities as functions of the horizon radius $r_+$. The inverse temperature has a simple expression as a function of $r_+$, namely
\begin{eqnarray}
\beta  &=& {4\pi \over \Psi_1'(r_+) } \\
&=& {2\pi l^2(4\alpha + r_+^2) \over r_+(l^2 + 2r_+^2) }\,, \label{beta}
\end{eqnarray}
and the mass was already given in (\ref{eqM}). Replacing (\ref{eqM}) and (\ref{beta}) into (\ref{dI}) we get
\begin{eqnarray}
\delta \tilde I &=& - {3\pi^2(2l^2\alpha + l^2r_+^2 +r_+^4)(4l^2\alpha - (l^2-24\alpha)r_+^2 + 2 r_+^2)  \over 4r_+^2(l^2+2r_+^2) } \, \delta r_+\,.
\end{eqnarray} 
This can be integrated explicitly to give the Euclidean action, 
\begin{eqnarray}
\tilde I = {\pi^2(\, 24l^2\alpha^2- 6l^2 \alpha r_+ +(l^2-36\alpha)r_+^4 - r_+^6 ) \over 4r_+(l^2 + 2 r_+^2) }\label{actionf}
\end{eqnarray} 
up to a constant. Nicely, the entropy has a very simple form, 
\begin{eqnarray}
S &=& -\tilde I + \beta M \\
&=& {2\pi^2 r_+^3 \over 4} + 6\pi^2\alpha r_+
\end{eqnarray} 
exhibiting the well-known $A/ 4$ (the area of the unit 3-shere is $2\pi^2$) plus a linear piece in $r_+$. The extra term is characteristic of Gauss-Bonnet interactions \cite{Jacobson:1993xs,Banados:1993ur}. Indeed, since $\alpha$ has dimensions length$^2$, only the combination $\alpha r_+$ can appear. 

The expressions for the temperature (\ref{beta}) and the action (\ref{actionf}) deserve some comments. First, (\ref{beta}) shows that the Gauss-Bonnet coupling $\alpha$ should be regarded as a positive parameters. If $\alpha$ was negative, small black holes would have an unphysical negative $\beta$. We consider here only $\alpha>0$ (at $\alpha=0$ we are back at standard 5d Schwarzschild AdS case). Now, in contrast with the Schwarzschild-AdS black hole, in this system the inverse temperature (\ref{beta}) ranges from zero to infinity. This means that black holes exists with all temperatures. It also follows that for any given temperature there exists a unique value of $r_+$, and then a unique value of $M$. The action on the other hand does have a sign flip when changing then temperature. For large temperatures $T=1/\beta$ (this is the same as $r_+\rightarrow\infty$), $\tilde I<0$, that is black holes have less action than the AdS groundstate and dominate. For small temperature ($r_+ \rightarrow 0$),  $\tilde I>0$;  the AdS background has less action becoming the preferred configuration.

\section{Conclusions}\label{sec:conclusions}
The Gauss-Bonnet invariant is a boundary term in 4 dimensions when the traceless part of nonmetricity is zero~\cite{Janssen:2019uao}. In higher dimensions, however, this term is no longer a boundary term and therefore acquires dynamical significance. Leveraging this fact, we formulated a Gauss-Bonnet theory within the framework of Weyl geometry, where nonmetricity is attributed solely to its Weyl component. In this context, the proposed Gauss-Bonnet theory represents the simplest way to incorporate nonmetricity.

We discovered that this theory admits an exact spherically symmetric black hole solution in 5 dimensions, characterized by a single horizon and a non-zero Weyl vector that contributes to the metric sector. Additionally, we demonstrated that it is possible to define the thermodynamics of this solution by ensuring that all fields (the metric and the Weyl vector) remain regular at the horizon. This approach enabled us to calculate the entropy of the black hole, which includes an additional contribution arising from the non-Riemannian Gauss-Bonnet coupling constant.

It would also be interesting to conduct a similar analysis to~\cite{Glavan:2019inb} to construct a 4-dimensional Gauss-Bonnet theory in the context of Weyl geometry. As shown in several studies~\cite{Lu:2020iav, Hennigar:2020lsl, Fernandes:2020nbq}, it is anticipated that such a theory would belong to a broader class of Horndeski gravity, with the Weyl component of nonmetricity playing a significant role. Furthermore, extending this framework to incorporate additional sources, such as electromagnetic fields or scalar potentials, could provide a richer class of solutions and shed light on the role of nonmetricity in coupled systems. On the other hand, it might be also interesting to include torsion and construct a Weyl-Cartan extension of this theory. All of these studies will be carried out in the future.

\bigskip
\bigskip
\noindent
\section*{Acknowledgements}

S.B. is supported by “Agencia Nacional de Investigación y Desarrollo” (ANID), Grant “Becas Chile postdoctorado
al extranjero” No. 74220006. The work of MB was supported by a UC-Chile VRA grant (apoyo a sabáticos). 

\newpage

\bibliographystyle{utphys}
\bibliography{references}

\end{document}